\documentclass[]{mn2e}
\usepackage{psfig, epsf, epsfig}
\title[Intracluster globular clusters]{Formation of
 intracluster globular clusters}
\author[Hideki Yahagi and Kenji Bekki]
       {Hideki Yahagi${}^1$\thanks{E-mail: hyahagi@astron.s.u-tokyo.ac.jp} 
        and Kenji Bekki${}^2$\thanks{E-mail: bekki@bat.phys.unsw.edu.au}\\
        ${}^1$Department of Astronomy, University of Tokyo,
              7-3-1 Hongo, Bunkyo ward, Tokyo 113-0033, Japan\\
        ${}^2$School of Physics, University of New South Wales,
              Sydney 2052, NSW, Australia\\}
  
\begin{document}

\date{Accepted, Received 2005 May 13; in original form }

\pagerange{\pageref{firstpage}--\pageref{lastpage}} \pubyear{2005}

\maketitle

\label{firstpage}

\begin{abstract}

We first present the results of numerical simulations 
on formation processes and
physical properties of old globular clusters (GCs) located
within clusters of galaxies (``intracluster GCs'') and in between clusters
of galaxies (``intercluster GCs'').
Our high-resolution cosmological
simulations with models of GC formation at high redshifts ($z>6$)
show  that  about 30 \% of all GCs in a rich cluster can be ragarded as 
intracluster GCs that can freely drift being trapped by
gravitational potential of the cluster  rather than by the cluster
member galaxies. 
The radial surface density profiles of 
the simulated intracluster GCs  are highly likely
to be flatter than those of GCs within  cluster member galaxies. 
We also find that  about 1\% of all GCs formed before $z>6$ are not located
within any virialized halos 
and can be regarded as   
``intercluster'' (or ``intergalactic'')
GCs.
We discuss the dependences of physical properties
of intracluster and intercluster GCs  on the initial density
profiles of GCs within low-mass dark matter halos at high redshifts ($z>6$).
\end{abstract}

\begin{keywords}
globular clusters: general --
galaxies: star clusters --
galaxies:evolution -- 
galaxies:stellar content
\end{keywords}

\section{Introduction}

Recent observational studies of globular clusters (GCs)
in clusters of galaxies have suggested that
there can be a population of GCs that are 
bounded by cluster gravitational potentials 
rather than those of cluster member
galaxies (e.g., West et al. 1995; Bassino et al. 2002, 2003; 
Jord\'an et al. 2003),
though the existence of these intracluster GCs (ICGCs) in the Coma cluster
is observationally suggested to be highly unlikely  
(e.g., Mar\'in-Franch \& Aparicio 2003).
Structural and kinematical studies of a population of
very bright star clusters $-$ known as ``ultra-compact dwarfs
(UCDs)'' $-$ have also suggested  that these clusters can be
also freely floating intracluster objects 
(Mieske et al. 2004; Drinkwater et al. 2005; Jones et al. 2005).
Physical properties of intracluster stellar objects 
such as ICGCs and PNe 
are considered to be sensitive to 
dark matter properties and
cluster-related physical processes (e.g., tidal stripping
of GCs and hierarchical growth of clusters)
and thus provide some fossil information on formation of
galaxies and clusters of galaxies 
(e.g., Arnaboldi 2004 for a recent review).

Although there have been developments on the observational front,
there has been little theoretical and numerical works  carried out 
as to how ICGCs are formed in clusters environments 
(e.g., Forte et al. 1982; Muzzio et al. 1987; Bekki et al. 2003). 
These previous models showed that 
tidal stripping of GCs from cluster member galaxies
though galaxy-galaxy and galaxy-cluster interaction
is a mechanism for ICGC formation.
These previous works however used {\it fixed} gravitational potentials of 
already virialized clusters 
and accordingly could not discuss {\it how ICGCs in  a cluster  are formed as
the cluster grows through hierarchical merging of smaller groups
and clusters.}
Thus it remains unclear (1) how ICGC are formed 
under the currently
favored cold dark matter (CDM) theory of galaxy formation
and (2) what physical properties  ICGCs can have if their formation
is closely associated with hierarchical formation of clusters.

\begin{figure*}
\psfig{file=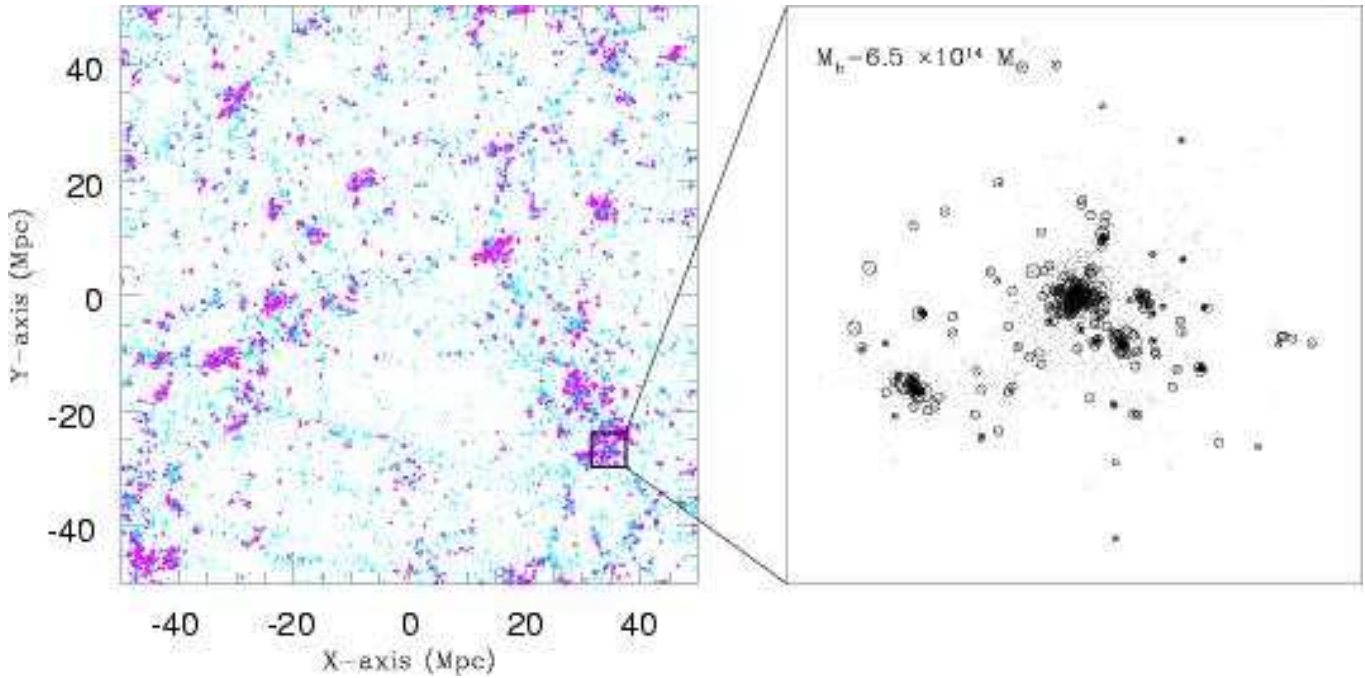,width=18.0cm}
\caption{ 
{\it Left}: The large scale distributions of INGCs (magenta big dots)
and HGCs (cyan dots)
projected onto the $x$-$y$ plane at $z=0$ in the fiducial model.
Here INGCs represent GCs that are not within any virialized
halos at $z=0$ and thus include intergalactic
and intercluster GCs.
HGCs are GCs that are within virialized halos at $z=0$.
{\it Right}: Distributions of GCs projected onto the $x$-$y$ plane
at $z=0$ for a cluster-scale halo 
with the total mass of $6.5 \times 10^{14} {\rm M}_{\odot}$.
GCs within circles represent those within tidal radii of
galaxy-scale halos and the radii of the circles
represent the tidal radii.
GCs that are not within any circles are regarded as ICGCs.
}
\label{Figure. 1} 
\end{figure*}

The purpose of this Letter is thus to demonstrate, for the first time,
how ICGCs are formed during hierarchical formation of clusters 
of galaxies, 
based on high-resolution cosmological simulations that can
follow both hierarchical growth of clusters through merging of
smaller subhalos 
and dynamical evolution of old GCs.
We also discuss physical properties of GCs that were formed
within subhalos at $z>6$ yet are not within any virialized halos
at $z=0$: These GCs can be regarded as ``intergalactic''
(van den Bergh 1958) 
or ``intercluster'' GCs.
For convenience and clarity, 
GCs within any virialized halos 
with the masses larger than $3 \times 10^9 {\rm M}_{\odot}$
are refereed to as Halo Globular Clusters (HGCs)
and the meanings of these acronym are given in Table 1.

\section{The model}

We simulate the large scale structure of GCs  
in a $\Lambda$CDM Universe with ${\Omega} =0.3$, 
$\Lambda=0.7$, $H_{0}=70$ km $\rm s^{-1}$ ${\rm Mpc}^{-1}$,
and ${\sigma}_{8}=0.9$ 
by using the Adaptive Mesh Refinement $N-$body code developed
by Yahagi (2005) and Yahagi et al. (2004), 
which is a vectorized and parallelized version
of the code described in Yahagi \& Yoshii (2001).
We use $512^3$ collisionless dark matter (DM) particles in a simulation
with the box size of $70h^{-1}$Mpc and the total mass 
of $4.08 \times 10^{16} {\rm M}_{\odot}$. 
We start simulations at $z=41$ and follow it till $z=0$
in order to investigate physical properties
of old GCs outside and inside virialized dark matter halos at $z=0$. 
We used the COSMICS (Cosmological Initial Conditions and
Microwave Anisotropy Codes), which is a package
of fortran programs for generating Gaussian random initial
conditions for nonlinear structure formation simulations
(Bertschinger 1995, 2001).

The way of investigating GC properties is described as follows.
Firstly we select virialized dark matter subhalos at $z=z_{\rm form}$ 
by using the friends-of-friends (FoF) algorithm (Davis et al. 1985)
with a fixed linking length of 0.2 times the mean DM particle separation.
The minimum particle number $N_{\rm min}$ for halos is set to be 10.
For each individual virialized subhalo
with the half-mass radius of $R_{\rm h}$,
some fraction ($f_{\rm gc}$) of particles within $R_{\rm h}/3$ are labeled 
as ``GC'' particles. This procedure for defining GC particles
is based on the assumption that energy dissipation via radiative cooling
allows baryon to fall into the deepest potential well of dark-matter halos
and finally to be converted into GCs.
The value of the truncation radius ($R_{\rm tr,gc}$ =  $R_{\rm h}/3$)
is chosen, because the size 
of the very old GCs in
the Galactic GC system 
(i.e., the radius within which most Galactic old GCs are located)
is similar to $R_{\rm h}/3$ of the dark matter
halo in the dynamical model of the Galaxy (Bekki et al. 2005).
We assume that old, metal-poor globular cluster formation is truncated
after $z=z_{\rm form}$, because previous theoretical studies
demonstrated that such truncation of GC formation 
by some physical mechanisms (e.g., reionization) is necessary
for  explaining  the color bimodality of GCs,
very high specific frequency ($S_{\rm N}$)
in cluster Es,  and structural properties of the Galactic 
old stars and GCs (e.g., Beasley et al. 2002; Santos 2003;
Bekki 2005; Bekki \& Chiba 2005).

Secondly we follow the dynamical evolution of GC particles
formed before $z=z_{\rm form}$ till $z=0$ and thereby
derive locations ($(x,y,z)$) 
and velocities ($(v_{\rm x},v_{\rm y},v_{\rm z})$)
of GCs at $z=0$.
We then identify virialized halos at $z=0$ with the FoF algorithm
and investigate whether each of GCs is within the halos.
If a GC is found to be within a halo, the mass of the host halo
($M_{\rm h}$)
and the distance of the GC from the center of the halo 
($R_{\rm gc}$) are investigated.
If a GC is not in any halos, it is regarded as an INGC
and the distance ($R_{\rm nei}$) between the INGC and the nearest neighbor
halo and the mass of the halo ($M_{\rm h,nei}$) are
investigated.
If a GC is found to be within a cluster-size halo 
($M_{\rm h} > 10^{14} {\rm M}_{\odot}$), 
we investigate which galaxy-scale halo in the cluster-scale
halo contains the GC.
We examine local mass densities around particles in a cluster
and thereby select galaxy-scale halos that have high densities enough to
be identified as galaxy-scale halos (BY). 
If we find the GC within the tidal radius ($R_{\rm t}$) of 
one of galaxy-scale halos 
in the cluster-scale halo, 
it is regarded as a galactic GC (GGC):
Otherwise it is regarded as an ICGC. 
$R_{\rm t}$ is assumed to be the radius 
where the slope $\alpha$ in the GCS density profile of 
$\rho (r) \propto r^{\alpha}$ in a galaxy-scale halo 
becomes smaller  than -5 (i.e., much steeper than the outer
profile of the dark matter halo).

\begin{table}
\centering
\begin{minipage}{75mm}
\caption{Meaning of acronym}
\begin{tabular}{|l||l|} 
INGC &  INtergalactic GCs \\
ICGC &  IntraCluster GCs \\
GGC &  Galactic GCs \\
HGC &  Halo GCs \\
\end{tabular}  
\end{minipage}
\end{table}

Thus, the present simulations enable us to investigate
physical properties only for old  GCS
owing to the adopted assumptions of collisionless simulations:
Physical properties of metal-rich
GCs lately formed during secondary dissipative galaxy  merger events
at lower redshifts (e.g., Ashman \& Zepf 1992) can not 
be predicted by this study.
We present the results of the model with $z_{\rm form}=6$,
and the dependences of the results on  $z_{\rm form}$
will be given in Bekki \& Yahagi (2005, BY).
If $z_{\rm form}$ is closely associated with the completion
of cosmic reionization, $z_{\rm form}$ may well range
from 6 (Fan et al. 2003) 
to 20 (Kogut et al. 2003). 
Physical properties of hypothetical GC particles for
ICGCs in a rich cluster with 
$M_{\rm h} = 6.5 \times 10^{14} {\rm M}_{\odot}$ 
are described for 
the model with $f_{\rm gc}=0.2$ in which
the number ratio  of GC particles to all particles 
is $1.5 \times 10^{-3}$ at $z=0$.
We adopt $f_{\rm gc}=0.2$ so that typical subhalos at $z=6$ can
contain at least one GC particle. The present results does not
depend on $f_{\rm gc}$ at all as long as $f_{\rm gc}\ge 0.1$.
Physical properties of
ICGCs in groups and clusters with different masses
will be given in our forthcoming papers (BY).

We assume that the initial radial ($r$) profiles of GCSs ($\rho (r)$) in
subhalos at $z=6$ are the same as those of the simulated dark matter
halos that can have the universal ``NFW'' profiles
(Navarro, Frenk, \& White 1996) with 
$\rho (r) \propto r^{-3}$ in their outer parts.
The mean mass of subhalos at z=6 in the present simulations is roughly
$1.8 \times  10^{10} {\rm M}_{\odot}$,
which is similar to the total mass
of bright dwarf galaxies. 
Minniti et al. (1996) found that the projected ($R$) density
profiles of GCSs in dwarfs is approximated as $\rho (R) \propto R^{-2}$, which
is translated roughly as  $\rho (r) \propto r^{-3}$
by using a canonical conversion formula from
$\rho (R)$ into $\rho (r)$ (Binney \& Tremaine 1987).
Therefore, the above assumption on $\rho (r)$ can be regarded as reasonable.
Thus we mainly show the fiducial model with $\rho (r)$ similar
to the NFW profiles and $R_{\rm tr,gc}=R_{\rm h}/3$.

Although we base our GC models
on {\it observational results of GCSs at z=0},
we  can not confirm whether 
the above 
$\rho (r)$ and  $R_{\rm tr,gc}$
of the fiducial model are
really the most probable  (and the best)
for  GCSs {\it for low-mass subhalos at z=6 owing to the  lack of
observational studies of GCSs at high redshifts}.
Therefore we investigate how the numerical  results depend on
initial $\rho (r)$ and   $R_{\rm tr,gc}$ of subhalos 
at $z=6$.
Since the dependences on $\rho (r)$ and  $R_{\rm tr,gc}$
are given in details by BY,
we briefly describe the dependence on $R_{\rm tr,gc}$,
which is the most important dependence for physical
properties of GCSs at $z=0$ (BY).

\begin{figure}
\psfig{file=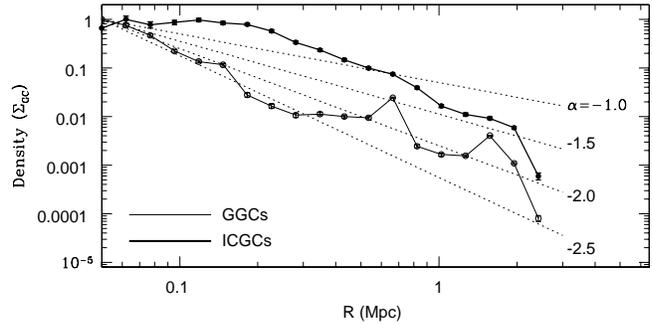,width=8.5cm}
\caption{ 
Projected radial density profiles (${\Sigma}_{\rm GC}(R)$) of
ICGCs (thick) and GGCs (thin).
Here GGCs represent ``galactic GCs'' that  
are GCs within any galaxy-scale halos in a cluster-scale halo.
Accordingly  GGCs are {\it not} GCs of the Milky Way:
We here use this term in order to distinguish  these GC populations 
(i.e., GGCs)  from
ICGCs. 
For clarity,
${\Sigma}_{\rm GC}(R)$ normalized to the maximum value
in each GC population is shown.
For comparison, the power-law density profiles with
${\Sigma}_{\rm GC}(R) \propto R^{\alpha}$ for
$\alpha$ = $-1.0$, $-1.5$, $-2.0$, and $-2.5$ are
shown by dotted lines.
}
\label{Figure. 2} 
\end{figure}

\section{Result}

\subsection{ICGCs}

Figure 1 shows 
the large scale ($\sim$ 100 Mpc) structure of INGCs
and HGCs
and  the distributions of GCs in a halo with
the total mass ($M_{\rm h}$) of 
$6.5 \times 10^{14} {\rm M}_{\odot}$ corresponding 
to a rich cluster of galaxies at $z=0$.
It is clear from Figure 1 that there exists ICGCs that
are not bounded in any cluster member galaxy-scale halos,
though most GCs  are within
the galaxy-scale halos.
About 29 \% of all GCs in this cluster
can be classified as ICGCs with the number fraction  of ICGCs 
ranging  from 0.28 at $R_{\rm cl} < 1$ Mpc 
(where $R_{\rm cl}$ is the distance between 
a GC and the center of the cluster)
to 0.35 at $1 \le  R_{\rm cl}  < 2$ Mpc.
Although the number fraction of ICGCs in the central 200 kpc
of the cluster is only 0.02,
the presence of such central ICGCs 
may well support  the scenario 
by West et al. (1995) that very high $S_{\rm N}$
of ICGCs in the central giant Es in some clusters can be due to
ICGCs.

Figure 2 shows that the projected number density distributions
(${\Sigma}_{\rm GC}$) within the central few hundreds kpc 
are quite different between ICGCs and GCs within
any galaxy-scale halos in the cluster (``galactic GCs'' 
referred to as ``GGCs'') 
in the sense that ${\Sigma}_{\rm GC}$ is significantly flatter
in ICGCs than in GGCs.
This is because most GCs in the central  few hundreds kpc of the cluster
can be identified as GCs within galaxy-scale halos
(i.e., smaller number of ICGCs). 
If GCs that are freely drifting under the influence of the cluster 
potential are located close to the central giant halo(s) of the cluster
(e.g., at their pericenter passages of orbital evolution),
they can be identified as GGCs in the present selection method
of ICGCs.
Therefore we suggest that 
${\Sigma}_{\rm GC}$ of ICGCs in   the central region of the cluster
can be somewhat underestimated in Figure 2.

The number fraction of ICGCs ranges from $\sim 0.2$ to $\sim 0.4$
for the simulated clusters with 
$1.0 \times 10^{14} {\rm M}_{\odot} \le
M_{\rm h} \le  6.5 \times 10^{14} {\rm M}_{\odot}$. 
The power-law slopes of ${\Sigma}_{\rm GC}$ range from $\approx -1.5$
to $\approx -2.5$ for the clusters with the above mass range.
About $10-20$\% of all GCs are located in the central 50 kpc of 
the simulated rich clusters, which implies that these inner GCs
formed at high redshifts ($z>6$)
can be responsible for high $S_{\rm N}$ of the central  cD galaxies.
The more details on the parameter dependences of ICGC properties
will be discussed in our forthcoming papers (BY).

\begin{figure}
\psfig{file=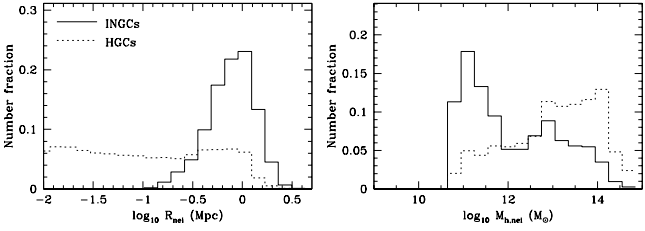,width=8.5cm}
\caption{ 
Distributions of $R_{\rm nei}$ (left) and $M_{\rm h,nei}$ (right)
for INGCs (solid) and for HGCs (dotted).
Here $R_{\rm nei}$ represents 
the distance between a GC and the center of a halo that is
nearest to the GC, and  
$M_{\rm h,nei}$ is the mass of the nearest neighbor halo.
Therefore, $R_{\rm nei}$ and $M_{\rm h,nei}$ for HGCs
are those of their host halos.
}
\label{Figure. 3} 
\end{figure}

\begin{figure}
\psfig{file=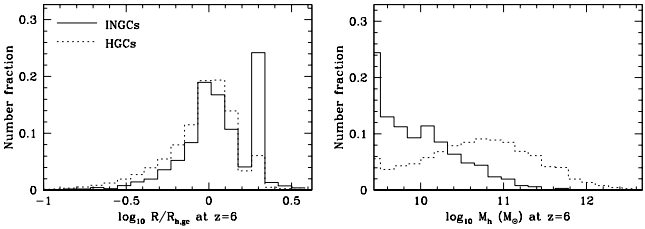,width=8.5cm}
\caption{ 
Distributions of $R/R_{\rm h,gc}$ (left) and $M_{\rm h}$ (right)
for INGCs (solid) and HGCs (dotted) at $z=6$.
Here $R$, $R_{\rm nei}$, and $M_{\rm h}$ represent
the distance of a GC from the center of its host halo
at $z=6$, the half-number radius of the GC system in the host halo,
and the mass of the host halo, respectively.
}
\label{Figure. 4} 
\end{figure}

\begin{figure}
\psfig{file=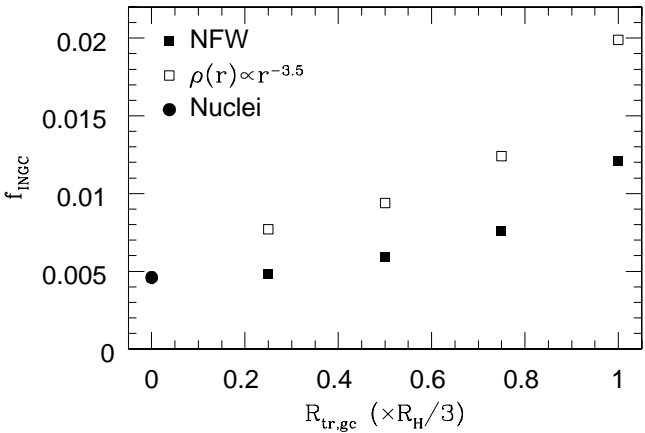,width=8.5cm}
\caption{ 
Dependences of the number fraction of INGCs ($f_{\rm INGC}$)
on the truncation radii of GCSs at $z=6$ ($R_{\rm tr,gc}$)
for the models with initial GC profiles similar to
density profiles of dark matter halos (filled squares) 
and power-law ones with the slope of $-3.5$ (open squares).
For comparison, the model in which only the very central
particle in each subhalo is regarded as a GC (represented by
``nuclei'') is shown by a filled circle. 
}
\label{Figure. 5} 
\end{figure}

\subsection{INGCs}

INGCs in the fiducial model 
comprises about 1\% of all GCs  formed in subhalos before $z=6$
(See Figure 1).
Although these INGCs are formed as a result of tidal stripping
of GCs from subhalos during hierarchical structure
formation through interaction and merging of subhalos between
$z=0$ and $z=6$, there 
appears to be no remarkable differences
in the large scale distributions between INGCs and  HGCs at $z=0$
(See Figure 1).
Figure 3 however indicates  clear differences in
$R_{\rm nei}$ and $M_{\rm h, nei}$  distributions
between INGCs and HGCs.
For example, the number fraction of INGCs with $R_{\rm nei}>1$ Mpc is 0.46 
whereas that of HGCs  with $R_{\rm nei}>1$ Mpc is 0.09.
The derived higher fraction in INGCs   
strongly suggests that INGCs are 
truly ``free-floating'' GCs in intergalactic/group/cluster regions.
The $M_{\rm h, nei}$ distribution of INGCs shows  two peaks 
around $M_{\rm h, nei} \approx 10^{11} {\rm M}_{\odot}$
and $\approx 10^{13} {\rm M}_{\odot}$ 
and the number fraction of $M_{\rm h, nei} > 10^{13} {\rm M}_{\odot}$ 
is 0.27 for INGCs, which is significantly smaller
than that (0.61)  for HGCs.
These results imply that
the fraction of intercluster GCs among all
INGCs can be observed to be small.

Figure 4 describes (1) where progenitor GCs of INGCs were located with respect
to the centers of their host subhalos at $z=6$ and
(2) what the mass ranges of their host subhalos were at $z=6$. 
The number fraction of GCs in the outer parts ($R/R_{\rm h,gc} >2$)
of their host subhalos at  $z=6$
is 0.17 for INGCs and 0.06 for HGCs in Figure 4.
This result suggests that GCs in the outer parts of subhalos,
where GCs are more strongly influenced by external tidal force,
are more likely to become INGCs and thus confirms that
tidal stripping of GCs during interaction and  merging of
subhalos between $z=0$ and $z=6$ is a  major mechanism for INGC formation. 
The mean masses  of host subhalos of INGCs and HGCs at $z=6$
are $2.5 \times 10^{10} {\rm M}_{\odot}$ and
$1.8 \times 10^{11} {\rm M}_{\odot}$, respectively, in Figure 4.
This is also consistent with the above formation process of INGCs,
because less massive subhalos are more strongly influenced by
tidal stripping.
If there are negative metallicity gradients of GC systems
and positive relations between halo masses and GC mean metallicities
in subhalos at $z=6$, 
as observed  for  nearby GC systems 
(e.g., Ostrov et al. 1993; C\^ote et al. 1998),
the results shown in Figure 4 imply that
INGCs can be significantly more metal-poor than HGCs.

Figure 5 describes how the number fraction of INGCs among
all GCs ($f_{\rm INGC}$) 
depends on initial density profiles of GCSs at $z=6$.
It is clear from this figure that (1) $f_{\rm INGC}$ is higher 
for larger $R_{\rm tr,gc}$ and (2) this $R_{\rm tr,gc}$ 
dependence can be
seen both in the NFW profile and the power-law one with
the slope of $-3.5$ (i.e., the observed profile of
the Galactic GCS). Figure 5 also shows that
only $\sim 0.5$\% of GCs that were initially in
the nuclear regions of subhalos at $z=6$ can finally
become INGCs.
These results imply that $f_{\rm INGC}$
can range from $\sim 0.005$ to $\sim 0.02$
for a reasonable set of model parameters on initial
density profiles of GCSs.

 $f_{\rm INGC}$ also depends on
the methods to identify halos and GCs. $f_{\rm INGC}$
is 0.0121, 0.0075, 0.0043, and 0.0024 for 
$N_{\rm min}$ =10, 20, 50, and 100, respectively.
The number fraction of INGCs that were nuclei at $z=6$ 
($f_{\rm INGC,N}$) are 0.0046, 0.0014, 0.0003, and 0.0001 for
$N_{\rm min}$ =10, 20, 50, and 100, respectively.
This $f_{\rm INGC,N}$ might well depend on the resolution
of simulations. 
$f_{\rm INGC}$ is 0.0067 in the model
with the FoF linking length of 0.025.
About 80 \% of GCs that are identified as INGCs
are {\it not bounded gravitationally} by any closest halos.  
These dependences imply  that $f_{\rm INGC}$ can range
from an order of 0.1\% to 1\%, given some uncertainties in
the best possible parameter values  of the methods
(e.g., FoF linking length).

\section{Discussions and conclusions}

Although van den Bergh (1958) already  suggested the existence
of INGCs in the local universe,
the present study first showed that (1) INGCs can be formed 
during hierarchical formation  processes of galaxies and clusters,
(2) these INGCs are about 1 \% of all GCs formed
before $z=6$, and (3) they can be $typically$ more metal-poor than those
within virialized galaxy-scale halos at $z=0$.
These INGCs are highly unlikely to suffer destruction processes
by strong galactic tidal fields that are suggested
to be important for understanding the origin 
of the observed mass function of GCs (e.g., Fall \& Zhang 2001).
Therefore, we suggest that (1) the mass function of INGCs 
can be significantly
different from that of GCs in galaxies and (2) INGCs possibly
retain fossil information on GC mass function  
at the epoch of GC formation.   

Then how many INGCs are expected to be observed in the intergalactic
regions close to the Galaxy ? 
We can provide an answer for this question by using the present 
result on the number fraction of INGCs and 
the initial GC number of the Galactic GCS
before GC destruction.
McLaughlin (1999) showed that total number of initial GCs
in a galaxy
can decrease by a factor of 25 within the Hubble time  
owing to GC destruction by the combination effect of  
galactic tidal fields and internal GC evolution
(e.g., mass loss from massive and evolved stars).
This means that the initial GC number 
is about  4000 
for the Galaxy with 
the observed GC number of 160
at $z=0$ (van den Bergh 2000).
By using the present result that $\sim 1$\% of all GCs 
can become INGCs,
the expected  number  of ICGCs in the intergalactic regions close to
the Galaxy can be  $\sim 40$.
We thus suggest that some bright objects of
these $\sim 40$ intergalactic GCs 
can be found in
currently ongoing ``all-objects'' spectroscopic surveys
for targeted areas (e.g., 6dF).

Since White (1987) pointed out  that clusters of galaxies
might contain ICGCs,  several authors have suggested 
some observational evidences for or against the existence
of ICGCs (e.g., West et al. 1995; Blakeslee 1997; Harris et al. 1998).
Our future more extensive numerical studies 
of ICGC formation 
will provide observable
predictions on some correlations between number of ICGCs and
global properties (e.g., mass and X-ray temperature of hot gas) in
their host clusters and thus help observers to confirm the 
existence of ICGCs.
Furthermore, if physical properties of ICGCs strongly depend on
$z_{\rm form}$
after which GC formation was severely suppressed
by some  physical processes (e.g., cosmic reionization),
they can provide some observational constraints
on $z_{\rm form}$.

\section*{Acknowledgments}
We are  grateful to the anonymous referee for valuable comments,
which contribute to improve the present paper.
The numerical simulations reported here were carried out on 
Fujitsu-made vector parallel processors VPP5000
kindly made available by the Astronomical Data Analysis
Center (ADAC) at National Astronomical Observatory of Japan (NAOJ)
for our  research project why36b.
H.Y. acknowledges the support of the research fellowships of the Japan
Society for the Promotion of Science for Young Scientists (17-10511).
K.B. acknowledges the financial support of the Australian Research 
Council throughout the course of this work.

\end{document}